\UseRawInputEncoding
\pdfoutput=1
\documentclass[aps,prl,reprint,showpacs,nolongbibliography]{revtex4-1}
\usepackage{amsmath}
\usepackage{graphicx}
\usepackage{hyperref}
\usepackage[all]{hypcap}
\usepackage{microtype}
\usepackage{changes}
\usepackage{bibentry}

\usepackage{array}
\usepackage{supertabular}
\usepackage{hhline}
\usepackage{multirow}
\usepackage{float}
\usepackage{booktabs}
\usepackage{tabularx}
\usepackage{xcolor}

\hypersetup{
    colorlinks,
    linkcolor={blue},
    citecolor={blue},
    urlcolor={blue}
}

\begin{document}
%\UseRawInputEncoding

\title{Assessing the potential of perfect screw dislocations in SiC for solid-state quantum technologies}

\author{Daniel Barragan-Yani and Ludger Wirtz}
\affiliation{Physics and Materials Science Research Unit, University of Luxembourg, 162a Avenue de la Fa{\"i}ncerie, L-1511 Luxembourg, Luxembourg}
%\author{Ludger Wirtz}
%\affiliation{Physics and Materials Science Research Unit, University of Luxembourg, 162a Avenue de la Fa{\"i}ncerie, L-1511 Luxembourg, Luxembourg}

\date{\today}

\begin{abstract}

{ Although point defects in solids are one of the most promising physical systems to build functioning qubits, it remains challenging to position them in a deterministic array and to integrate them into large networks. By means of advanced ab initio calculations we show that undissociated screw dislocations in cubic 3C-SiC, and their associated strain fields, could be used to create a deterministic pattern of relevant point defects. Specifically, we present a detailed analysis of the formation energies and electronic structure of the divacancy in 3C-SiC when located in the vicinity of this type of dislocations. Our results show that the divacancy is strongly attracted towards specific and equivalent sites inside the core of the screw dislocations, and would form a one-dimensional arrays along them. Furthermore, we show that the same strain that attracts the divacancy allows the modulation of the position of its electronic states and of its charge transition levels. In the case of the neutral divacancy, we find that these modulations result in the loss of its potential as a qubit. However, these same modulations could transform defects with no potential as qubits when located in bulk, into promising defects when located inside the core of the screw dislocations. Since dislocations are still mostly perceived as harmful defects, our findings represent a technological leap as they show that dislocations can be used as active building blocks in future defect-based quantum computers.
}

\end{abstract}


\maketitle

A quantum computer is a device that exploits quantum behavior to deal with computational problems that cannot be solved, or would take too long, in a classical computer~\cite{nielsen2010quantum, Ladd2010}.  In order to build a functioning quantum computer, several physical systems have been proposed as platforms for quantum bits (qubits), e.g., photons~\cite{Flamini_2019, doi:10.1063/1.5115814}, trapped atoms/ions~\cite{HAFFNER2008155, doi:10.1063/1.5088164} and point defects in solids~\cite{doi:10.1126/science.276.5321.2012, Balasubramanian2009, DOHERTY20131, doi:10.1063/5.0006075, Chatterjee2021, Wolfowicz2021}. The latter system is advantageous from the point of view of scalability since integrated quantum devices could, in principle, be built by means of adapted fabrication techniques developed in the semiconductor industry~\cite{koehl_seo_galli_awschalom_2015}. Nevertheless, it remains challenging to position the point defects in a deterministic array and to integrate them into large networks~\cite{Atature2018}.

Currently, the desired array of point defects are created using mainly irradiation and implantation. However, these methods are limited by the beam size on target (positioning accuracy of $\geq$ 30 nm, when achieving a high creation yield)~\cite{Schröder2017, PhysRevApplied.7.064021, Zhou_2018, Chen:19, SmithMeynellBleszynskiJayichMeijer, Castelletto_2020_REVIEW}. As has been pointed out for the case of two-dimensional materials, alternative to more traditional methods, an engineered strain field is also able to create an array of point defects used as quantum emitters~\cite{Kumar2015, Branny2017}. Nevertheless, in that case the regularity of the resulting array is hampered by the precision of the method used to grow the nanopillars that induce the strain field (positioning accuracy of 120$\pm$32 nm).

Besides the localization and interconnection problems, one further limitation of current defect-based quantum technologies is that usually only point defects are considered as active elements. As a matter of fact, in their seminal work on the potential of point defects for quantum computing, Weber et al. proposed that one of the properties of a suitable host material would be its availability as high-quality single crystals~\cite{Weber2010Defects}. Up to now, the presence and effects of extended defects (e.g., dislocations, stacking faults, grain boundaries and surfaces) has been considered mostly from the point of view of irregularities to avoid and control~\cite{Wolfowicz2021, Stamp-Gaita, Alkauskas_Design, Kaviani2014, chou_gali_2017, Chou2017, Stacey2019, PhysRevX.9.031052, LI2019273, Shen2020, PhysRevB.103.085305, PhysRevB.105.085305, doi:10.1063/5.0080096, PRXQuantum.3.040328}.  It was only recently that there has been interest on understanding the possible advantages and applications of extended defects, specifically stacking faults~\cite{Ivády2019}.

\begin{figure*}[t]
%\vspace{-0.3cm}
%\hspace{-1cm}
\includegraphics[scale=0.65]{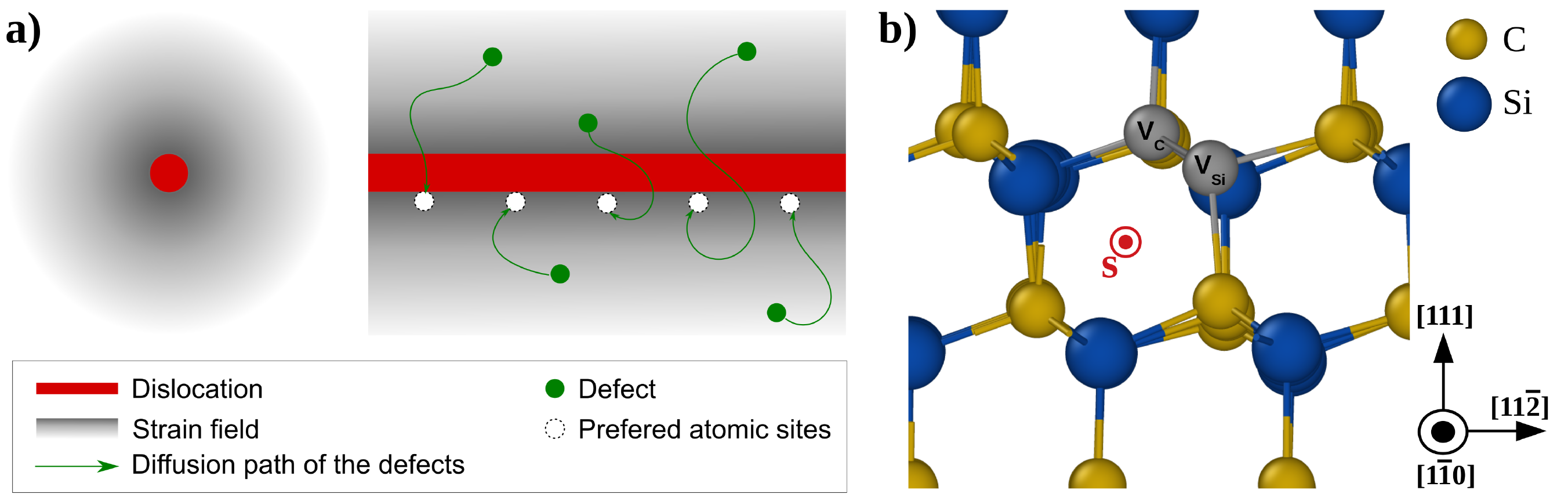} 
\vspace{-0.2cm}
\caption{\label{f:estructuras} (a) Depiction of a dislocation (red line) attracting defect-based qubits (green dots) due to its induced strain field (gray shaded area). (b) Example of the preferred configuration of the DV0 when located near a screw dislocation in 3C-SiC, i.e., both vacancies located directly at the core of the dislocation. Figure (b) was obtained using OVITO~\cite{0965-0393-18-1-015012}.}
\end{figure*}

Compared to stacking faults, dislocations have a stronger and longer-range interaction with point defects due to their characteristic strain fields~\cite{holt2007extended, hull2011introduction, cai2016imperfections, anderson2017theory}. This means that, as illustrated in Fig.~\ref{f:estructuras}(a), qubits built using point defects could be created near an engineered dislocation and, due to the attraction between the qubits and the dislocation, an annealing process would result in a self-assembled one-dimensional array of qubits along the dislocation. Recently, it was shown that certain positions inside the cores of partial dislocations in diamond are preferred locations for NV$^-$ centers to form, while conserving the interesting quantum properties of the centers~\cite{PhysRevB.106.174111}. However, partial dislocations are not radially symmetric, and several undesired configurations of the NV$^-$ centers near partial dislocations in diamond are likely to exist~\cite{PhysRevB.106.174111}. One possible way to overcome these limitations and to effectively harness the potential of dislocations for quantum technologies would be to engineer perfect (undissociated) dislocations, specifically screw dislocations, which have a radially symmetric strain fields and core structures~\cite{holt2007extended, hull2011introduction, cai2016imperfections, anderson2017theory}. Another advantage of using this type of dislocations is the fact that it is possible to create ordered and regular two-dimensional arrays of perfect screw dislocations using wafer bonding techniques~\cite{Reiche_2010, Reiche_2014}, which would imply the possibility of creating self-assembled two-dimensional arrays of qubits along the screw dislocations.

For this idea to work, we need host materials which satisfy two conditions: screw dislocations prefer to be undissociated and, in that preferred configuration, they are electrically inactive. The latter condition ensures that the relevant quantum properties of the defect-based qubits located along the dislocation are not altered beyond the effect of the strain field associated with the screw dislocation. Among the most relevant host materials for defect-based quantum technologies, silicon carbide is the one exhibiting perfect screw dislocations as stable configurations under normal conditions~\cite{Pizzagalli_2005, https://doi.org/10.1002/pssc.201200372, PIZZAGALLI2014236}. 

In this Letter, we use ab initio simulations to study the strength and effects of the interaction between a perfect screw dislocation in 3C-SiC and the divacancy (DV) in the same material. The DV in 3C-SiC is known to exhibit promising properties as a qubit when in its charge neutral (DV0) configuration~\cite{doi:10.1063/5.0006075, Castelletto_2020_REVIEW, PhysRevX.7.021046, doi:10.1063/5.0004454}. To carry out our study, we analyze the change in the formation energy and electronic structure of the DV0 when located near the core of the screw dislocation. Our calculations reveal that DV0 prefers to form inside the core at specific and equivalent sites, implying that a one-dimensional array of qubits along the dislocation would be created. In addition, our results show that the same strain that drives the DV0 towards the dislocation core allows the modulation of the position of its electronic states and of its thermodynamic charge transition levels (CTLs).

Perfect screw dislocations in 3C-SiC are located in the shuffle set of the {111} planes with both\added{,} its dislocation line and Burgers vector lying in the $\langle$110$\rangle$ directions~\cite{Pizzagalli_2005, https://doi.org/10.1002/pssc.201200372, PIZZAGALLI2014236}. To study such dislocations in their stoichiometric and DV decorated configurations, the structural model used in our simulations contains a dislocation dipole in a quadrupolar arrangement, which minimizes the interaction between dislocations~\cite{bulatov2006computer, RODNEY2017633, Clouet2020}. Such model is created in a 576-atom SiC supercell using the software package $\mathsf{BABEL}$~\cite{babel}. This configuration can be seen in the supplementary material, Fig.~\ref{f:config_dipole}, and guarantees a minimum distance between dislocation cores of $\sim$16\AA . Our density functional theory (DFT) calculations are performed using the $\mathsf{VASP}$~\cite{PhysRevB.54.11169} simulation package with a converged plane-wave energy cutoff of 500 eV. The used supercells are built with the experimental lattice parameter of 4.36\AA ~\cite{Li1986} ~and ions are relaxed using the Perdew-Burke-Enzerhof (PBE)~\cite{GGA} semi-local functional until forces are below 0.02 eV/\AA . To obtain accurate formation energies, charge transition levels and electronic properties, we use the Heyd, Scuseria and Enzerhof (HSE06)~\cite{HSE06_1,HSE06_2} hybrid functional on the PBE relaxed supercells. In all cases we use a converged $2 \times 2 \times 2$ \textit{k}-point grid. To obtain a reference for our simulations, we performed calculations for the isolated DV in an analogous supercell and using the same computational setup.

Using the described methodology we tackle the first question at hand: \textit{does the DV prefer to form in the vicinity of perfect screw dislocation in 3c-SiC?} To answer it, we compare the formation energy in a given charge state $q$ of an isolated DV, $E^{\rm f}_{\rm iso}[DV^{q}]$, and that of the same defect when located near the core of the studied dislocation, $E^{\rm f}_{\rm dislo}[DV^{q}]$, which are defined as~\cite{RevModPhys.86.253, PhysRevB.86.045112}
\begin{multline}\label{e:formation_energy_iso}
E^{\rm f}_{\rm iso}[DV^{q}] = E^{\rm tot}_{\rm iso}[DV^{q}] - E^{\rm tot}[\rm bulk] - \sum_i n_i \mu_i    \\
  + q[E_{\rm F}+\epsilon_{\rm VBM}] + E_{\rm corr}^{q} ,
\end{multline}
and
\begin{multline}\label{e:formation_energy_dislo}
E^{\rm f}_{\rm dislo}[DV^{q}] = E^{\rm tot}[(DV+ \rm dislo)^{q}] - E^{\rm tot}[\rm dislo]     \\
  - \sum_i n_i \mu_i + q[E_{\rm F}+\epsilon_{\rm VBM}] + E_{\rm corr}^{q} ,
\end{multline}
where $E^{\rm tot}_{\rm iso}[DV^{q}]$ is the total energy of the supercell containing an isolated DV in charge state $q$, and $E^{\rm tot}[(DV+ \rm dislo)^{q}]$ is the total energy of the supercell containing the dislocation dipole with one DV, in charge state $q$, near one of the cores. The second term on the right side of both Eqs.~\ref{e:formation_energy_iso} and~\ref{e:formation_energy_dislo} is the total energy of the corresponding reference supercell. In the case of the isolated DV, $E^{\rm tot}[\rm bulk]$ is the total energy of the pristine supercell. For the case of the DV decorated core, $E^{\rm tot}[\rm dislo]$ is the total energy of the supercell containing the stoichiometric dislocation dipole. The last three terms on the right in Eqs.~\ref{e:formation_energy_iso} and~\ref{e:formation_energy_dislo} are analogous in both cases. Specifically, in the third term $n_i$ is an integer that indicates the number of atoms of type $i$ removed to create the DV and $\mu_i$ is the chemical potential of the same removed atoms. The fourth term, $q[E_{\rm F}+\epsilon_{\rm VBM}]$, accounts for the energy needed to charge the system. It contains the valence band maximum (VBM) energy, $E_{\rm VBM}$, and the Fermi energy, $E_{\rm F}$, referenced to the VBM (i.e., their sum $E_{\rm VBM} + E_{\rm F}$ tell us where, inside the band gap, is the chemical potential of the electrons). The last term, $E_{\rm corr}$, accounts for the correction needed to eliminate the spurious electrostatic interactions between periodic images of the defects under study. This correction is calculated using the approach proposed by Freysoldt, Neugebauer and Van de Walle~\cite{PhysRevLett.102.016402}. Note that this approach is only applicable if the dislocations under study are electrically inactive and all the charge added to the supercells is localized at the DVs. To prove the reliability of our methodology, in the supplementary material (Fig.~\ref{f:charge_localization}) it is possible to see the localized nature of the charge density difference between a DV0 and a charged DV when located directly at the core of the dislocation. The charge density difference was obtained using VESTA~\cite{Momma:db5098}.         

Once $E^{\rm f}_{\rm iso}[DV^{q}]$ and $E^{\rm f}_{\rm dislo}[DV^{q}]$ are defined, we can give a quantitative estimation of the strength of the interaction between the screw dislocation and the DV by calculating  
%\vspace{-0.4cm}
\begin{equation}\label{e:delta}
\Delta E^{\rm f}[DV^{q}] = E^{\rm f}_{\rm dislo}[DV^{q}] - E^{\rm f}_{\rm iso}[DV^{q}],
\end{equation}
which is negative if the DV prefers to be located near/inside the dislocation core. In the case of the neutral DV0, we found that there is a clear and strong attraction towards the dislocation. Specifically, the DV0 exhibits a $\Delta E^{\rm f}[DV^{0}] = -4.2$ eV for all configurations that have both vacancies occupying sites directly at the core of the dislocation. An example of such arrangement is shown on Fig.~\ref{f:estructuras}(b). Our calculations show that any other configuration, with the DV0 located further away from the core, implies steep changes in the corresponding $\Delta E^{\rm f}[DV^{0}]$ and a weaker interaction between the DV0 and the dislocation, i.e., a less negative $\Delta E^{\rm f}[DV^{0}]$. For example, the second-preferred configuration has the silicon vacancy directly at the core and the carbon vacancy siting outside the core, and exhibits a $\Delta E^{\rm f}[DV^{0}] = -3.7$ eV. The third-preferred configuration, when both vacancies composing the DV0 are located directly outside the core and in the second ring from the center, has a $\Delta E^{\rm f}[DV^{0}] = -2.6$ eV. The difference of +0.5 eV between the $\Delta E^{\rm f}[DV^{0}]$ of the preferred and of the second-preferred, and of +1.6 eV between the $\Delta E^{\rm f}[DV^{0}]$ of the preferred and of the third-preferred configurations exemplify the strong thermodynamic driving force that pushes the DV0 towards a specific configuration along the dislocations, with both vacancies occupying sites directly at the core. Structural examples of the second- and third-preferred configurations can be seen in Fig.~\ref{f:otherConfigs} of the supplementary material.

\begin{figure}[t]
%\vspace{-0.3cm}
%\hspace{-1cm}
\includegraphics[scale=0.65]{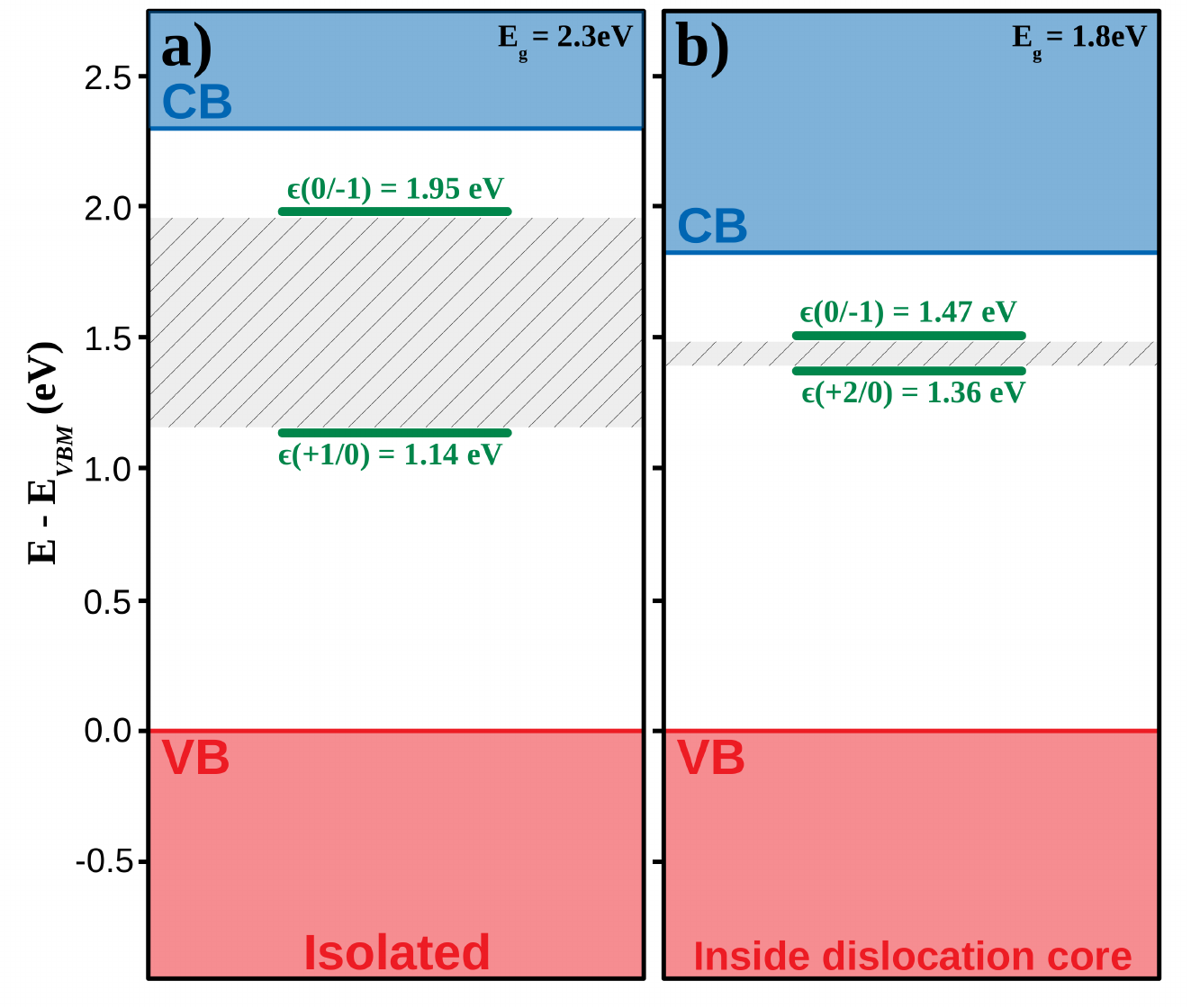} 
\vspace{-0.2cm}
\caption{\label{f:CTLs} Charge transition levels for the DV when (a) isolated and (b) located in its preferred configuration with both vacancies sitting directly at the core of the screw dislocation. In both figures, green horizontal lines represent represent the charge transition levels and dashed areas highlight the stability region for the neutral divacancy DV0. Band gaps for both cases are given on the top-right corner of each figure.}
\end{figure}

\begin{figure}[t]
%\vspace{-0.3cm}
%\hspace{-1cm}
\includegraphics[scale=0.65]{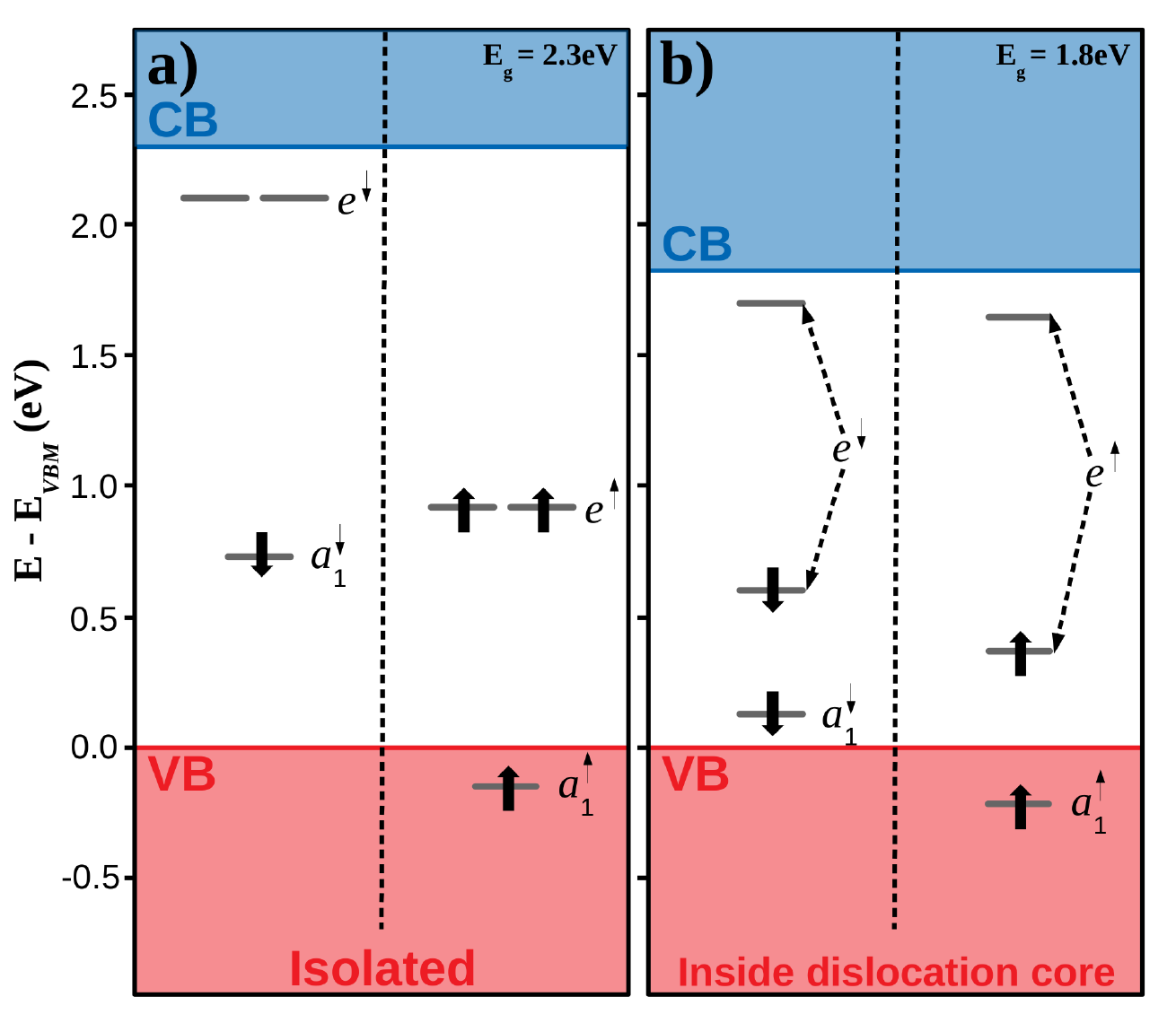} 
\vspace{-0.2cm}
\caption{\label{f:KS} Kohn-Sham states of the DV0 when (a) isolated and (b) located in its preferred configuration with both vacancies sitting directly at the core of the screw dislocation. In both figures, horizontal lines within the band gap point out the position of the obtained Kohn-Sham states and vertical arrows depict their corresponding occupations. Although the degeneracies of the $e^{\uparrow , \downarrow}$ states are lifted when the DVO is located inside the core of the screw dislocation, for simplicity we continue to use the same labels as in the isolated case. Band gaps for both cases are given on the top-right corner of each figure.}
\end{figure}

Although this analysis proves that the neutral DV0 is strongly attracted towards specific configurations directly at the core, a complete answer to the aforementioned question can only be given if other stable charge states are also addressed. For this purpose, we compare the CTLs of the isolated DV and those of the same defect when located near the dislocation. 
Besides allowing us to understand the electrical properties of defects (ionization energies), we decided to use the CTLs due to the fact that their position in the band gap provide us the stability range of the possible charge states.
For a given defect, the CTLs are referred to as $\epsilon (q/q')$ and are defined as the Fermi energy values at which its formation energy for charge states $q$ and $q'$ become equal~\cite{RevModPhys.86.253, PhysRevB.86.045112}. Our results for the $\epsilon (q/q')$ are shown in Fig.~\ref{f:CTLs}. On the one hand, in the case of the isolated DV, we find two CTLs, $\epsilon (+1/0) = 1.14$ eV and $\epsilon (0/-1) = 1.95$ eV. This means that in 3C-SiC, the DV0 is stable in a large Fermi energy region of $0.81$ eV for n-type doping conditions (around $35$ \% of the calculated $2.36$ eV band gap for bulk), which is in line with the corresponding results available in literature~\cite{PhysRevB.92.045208, PhysRevB.103.195202}. On the other hand, when the DV is located at its preferred configurations directly at the core of the screw dislocation, the stability region for the neutral DV0 is drastically reduced. In this case we found two CTLs, $\epsilon (+2/0) = 1.36$ eV and $\epsilon (0/-1) = 1.47$ eV, that imply a stability region of only $0.11$ eV, which corresponds to $\sim6$\% of the 1.8 eV band gap obtained for a dislocated sample (the reduced band gap observed for the supercell containing the dislocation dipole is caused by the band tail states associated to the disorder induced by the dislocations~\cite{PhysRevB.61.16033, PhysRevB.95.115203}). The fact that the stability region of the DV0 is reduced despite its strong interaction with the screw dislocation with $\Delta E^{\rm f}[DV^{0}] = -4.2$ eV, means that the interaction between the strain induced by the dislocation and the DV in all its charge states is of the same order of magnitude but larger for higher charges, which is common to other materials and defects~\cite{doi:10.1063/1.4958716, PhysRevB.95.195209}. Our calculations predict that the configuration preferred by the DV0, with both vacancies sitting directly at the core of the screw dislocation, is also the preferred one for its other stable charge states. Specifically, we obtained $\Delta E^{\rm f}[DV^{+2}] = -5.17$ eV and $\Delta E^{\rm f}[DV^{-1}] = -4.67$ eV. These findings imply that screw dislocations in 3C-SiC are able to attract defect-based qubits and induce the formation of one-dimensional qubit arrays, but for the specific case of the DV0 the stability range of the neutral state is drastically reduced. Nevertheless, the possibility of defects with higher charge states reacting more strongly to the strain induced by the dislocation, implies that new defect-based qubits, perhaps unstable in bulk, could be stabilized near screw dislocations in 3C-SiC.    

At this point in our assessment we need to answer a second question: \textit{are the electronic properties that make a given defect useful as a qubit, unaltered when located in the core of the screw dislocation?} Our analysis proved that, due to a reduced stability region, the DV0 would have a limited applicability when located inside the core of the screw dislocation. Nevertheless, we can study its Kohn-Sham states to reveal the effect of the strain field induced by the dislocation on the electronic structure of a potential qubit. In line with available literature on the matter~\cite{PhysRevB.92.045208, doi:10.1063/5.0080514, PhysRevMaterials.6.036201}, our calculations show that the isolated DV0 introduces an occupied $a^{\downarrow}_1$ state, a doubly occupied and degenerate $e^{\uparrow}$ state, and an unoccupied and degenerate $e^{\uparrow}$ state within the band gap of bulk 3C-SiC (see Fig.~\ref{f:KS}(a)). In addition, we also find an occupied $a^{\downarrow}_1$ state just below the valence band maximum. Altogether, the induced states imply a $S=1$ spin triplet ground state for the isolated DV, and its electronic structure satisfies the criteria required for defect-based qubits~\cite{Weber2010Defects}. As can be seen in Fig.~\ref{f:KS}(b), in the case of the DV0 located in its preferred configuration inside the screw dislocation core, the degeneracy of the $e^{\uparrow , \downarrow}$ states is lifted. The cause of this change is the localized disorder near the cores, which can be understood using a coordination analysis for the region near the core (see Fig.~\ref{f:Coord} of the supplementary material). This same disorder is the cause of the band tail states that reduce the band gap in the dislocated supercell~\cite{PhysRevB.61.16033, PhysRevB.95.115203}). The consequence of the degeneracy being lifted is that, compared to the isolated DV0, the DV0 near the core does not introduce a doubly occupied $e^{\uparrow}$ state, but a singly occupied $e^{\downarrow}$ state and a singly occupied $e^{\uparrow}$ state. Together with the occupied $a^{\uparrow , \downarrow}_1$, which are common to both the isolated and inside-the-core DV0, the electronic structure shown in Fig.~\ref{f:KS}(b) implies a $S=0$ spin singlet. This means that, besides seeing its stability region reduced, the DV0 located inside the core of the screw dislocation is not useful as a qubit. However, as it can also be seen in Fig.~\ref{f:KS}(b), the screw dislocation does not induce any deep state by itself and it allows a defect sitting in its core to have bound states. These findings imply that new defect-based qubits, perhaps useless in bulk, could become paramagnetic by having their electronic degeneracy being lifted near screw dislocations in 3C-SiC.      

In conclusion, we propose that, in order to have potential for quantum applications, dislocations should be undissociated screws and be electrically inactive. Such conditions are satisfied in 3C-SiC, and our calculations show that the undissociated screw dislocation in this material is able to attract defect-based qubits into its core. As a consequence, it would allow the creation of a one-dimension array of qubits along its line direction. Furthermore, we show that the strain field induced by this specific dislocation type is able to modulate both, the charge transition levels and the electronic structure of the qubit located in its core, without itself being electrically active. For the specific case of the DV0, our results show that these modulations result in the loss of its potential as a qubit. However, these same modulations could transform defects with no potential as qubits when located in bulk, into promising options when located inside the core of the screw dislocations. Altogether our findings represent a paradigm shift within quantum technologies, as they point out that dislocations can be used as active building blocks of future defect-based quantum computers.

We acknowledge that this project has received funding from the European Union's Horizon 2020 research and innovation programme under the Marie Sk{\l}odowska-Curie grant agreement No 898860. Furthermore, the authors gratefully acknowledge the computing time granted by the HPC facilities of the University of Luxembourg and the HRZ (Lichtenberg-Cluster) at TU
Darmstadt.

%{\color{purple} CONCLUSIONS Although in the case of the DV0 in SiC these modulations result in the lost of its potential as a qubit, the presence of... we need host materials where two conditions are satisfied: screw dislocations prefer to be undissociated and, in that preferred configuration, they are electrically inactive. The latter condition ensures that the relevant quantum properties of the defect-based qubits located along the dislocation are not altered beyond the effect of the strain field associated to the screw dislocation. } 

%\begin{figure}[b]
%\input{./coord} 
%\vspace{-0.2cm}
%\caption{\label{f:coord} Kohn-Sham states of the DV0 when (a) isolated and (b) located in its preferred configuration with both vacancies sitting directly at the core of the screw dislocation. In both figures, horizontal lines within the band gap point out the position of the obtained Kohn-Sham states and arrows depict their corresponding occupations. Band gaps for both cases are given on the top-right corner of each figure.}
%\end{figure} 

%\begin{itemize}
%\itemsep0em
%  \item[--] Reduced-size supercells, which cause artificially delocalized defect states.
%  \item[--] Incorrect description of the dielectric properties of the monolayers.
%  \item[--] Severe under- or overestimation of the band gap.
%  \item[--] Incorrect or lack of correction for the electrostatic interaction between charged defects and their periodic images.
%  \item[--] Incomplete assessment of the possible charge states of the defects.
%\end{itemize}

\section{Supplementary material}

\subsection{Supercell configuration}

\begin{figure*}[h]
%\vspace{-0.3cm}
%\hspace{-1cm}
\includegraphics[scale=0.65]{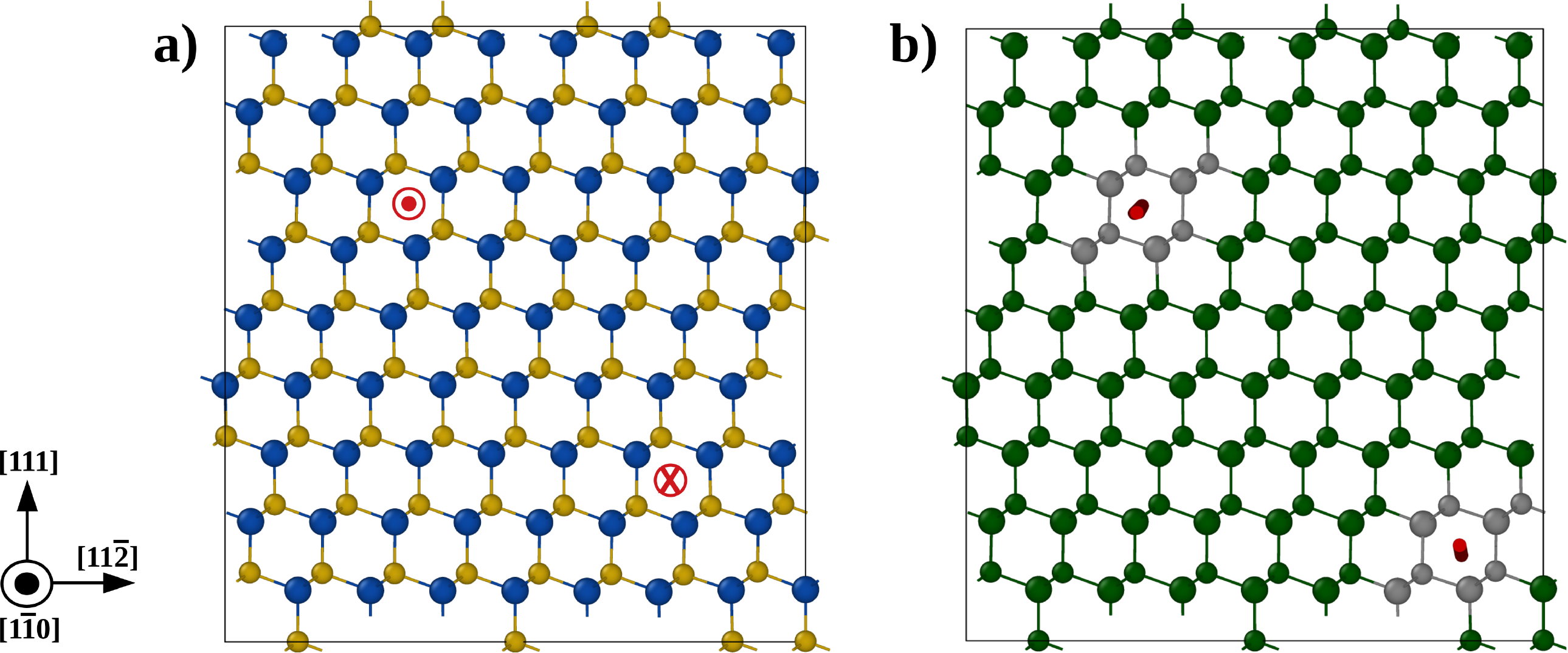} 
%\vspace{-0.4cm}
\caption{\label{f:config_dipole} (a) Screw dislocation dipole in a 576-atom supercell of 3C-SiC, with the position of dislocations and their line directions (in- and out-of-plane) depicted using vector notation (carbon and silicon atoms are shown in yellow and blue, respectively). (b) Dislocation analysis of the created supercells with dislocations shown as red lines, green atoms refer to bulk-like atomic sites and gray atoms signal atomic sites that cannot be clasify as diamond-like, i.e., the sites composing the core of the dislocation. Figures and analysis were obtained using OVITO~\cite{0965-0393-18-1-015012}.}
\end{figure*}

To study perfect stoichiometric and decorated screw dislocations in 3C-SiC, we used supercells containing a dislocation dipole in a quadrupolar arrangement, which minimizes the interaction between dislocations~\cite{bulatov2006computer, RODNEY2017633, Clouet2020}. Such structural model is created in a 576-atom SiC supercell using anisotropic elasticity via the software package BABEL~\cite{babel}. This configuration can be seen in Fig.~\ref{f:config_dipole}(a), and it guarantees a minimum distance between dislocation cores of $\sim$16\AA . In Fig.~\ref{f:config_dipole}(a), the position of dislocations and their line directions (in- and out-of-plane) are depicted using red marks that follow the usual vector notation. In Fig.~\ref{f:config_dipole}(b), we use the dislocation analysis tool available within OVITO~\cite{0965-0393-18-1-015012, Stukowski_2010} to detect and inspect the created screw dislocations within the supercell. Also in Fig.~\ref{f:config_dipole}(b), dislocation lines as extracted by OVITO are shown as red lines, green atoms refer to bulk-like atomic sites and gray atoms signal atomic sites that cannot be clasify as diamond-like, i.e., the sites composing the core of the dislocation.

\subsection{Charge localization}

\begin{figure*}[h]
%\vspace{-0.3cm}
%\hspace{-1cm}
\includegraphics[scale=0.65]{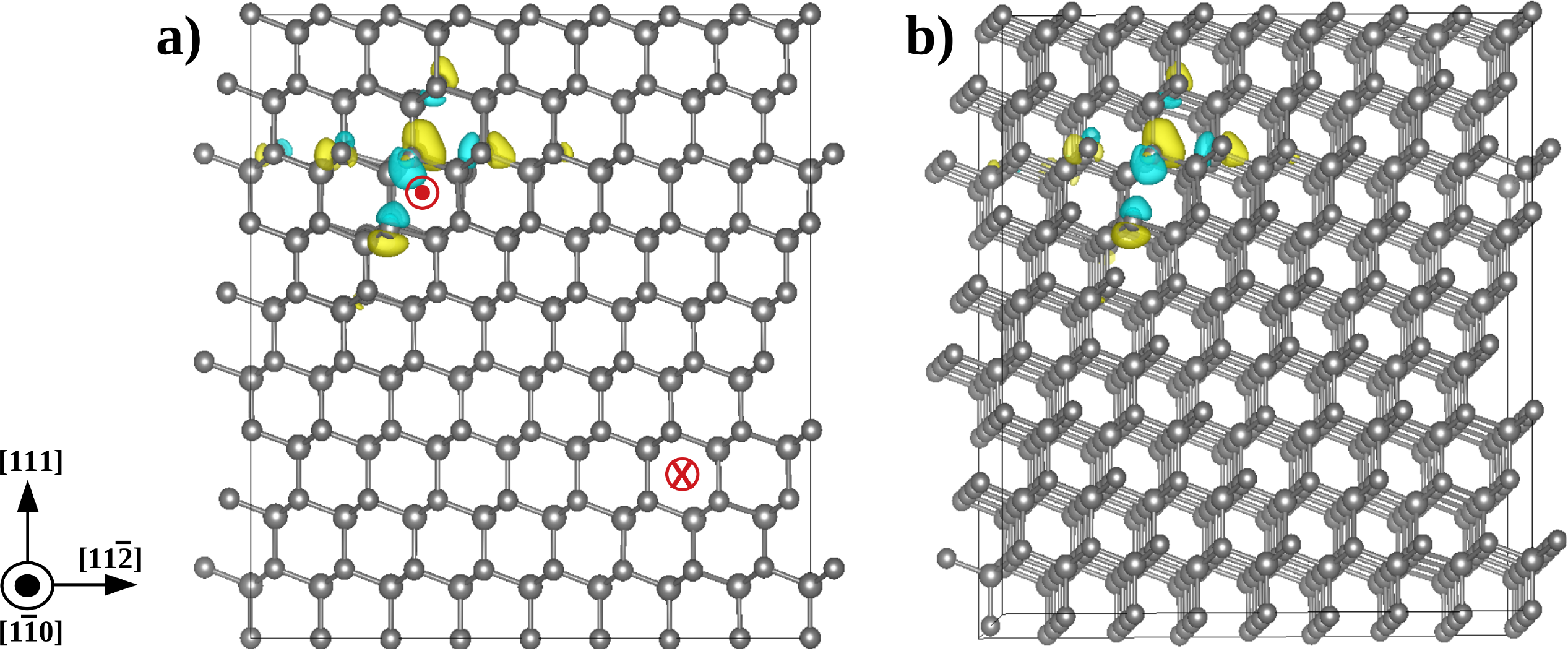} 
%\vspace{-0.4cm}
\caption{\label{f:charge_localization} (a) and (b) Two different perspectives of the charge density difference between a DV0 and a negatively charged DV when located directly at the core of one of the screw dislocations composing the dipole. Figures were obtained using VESTA~\cite{Momma:db5098}. The isosurfaces are shown at 5\% of the maximum.}
\end{figure*}

In Figs.~\ref{f:charge_localization}(a) and ~\ref{f:charge_localization}(b) we show two different perspectives of the charge density difference between a DV0 and a negatively charged DV when located directly at the core of the dislocation. As can be seen there, the additional charge is localized near the DV and it is not associated to any of the cores composing the dislocation dipole. This localization is a key requirement for the validity of our formation energy calculations and it confirms the electrical inactivity of the screw dislocations in 3C-SiC, which is one of the conditions that need to be satisfied for dislocations to be useful for quantum applications.

\subsection{Second- and third-preferred configurations}

\begin{figure*}[h]
%\vspace{-0.3cm}
%\hspace{-1cm}
\includegraphics[scale=0.65]{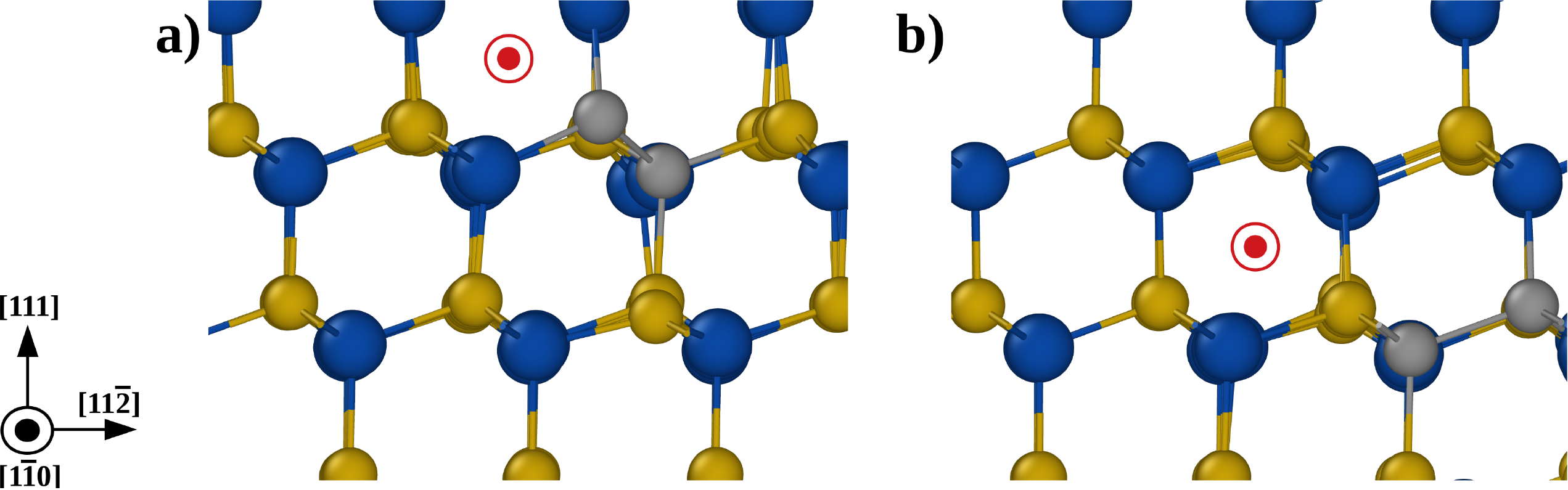} 
%\vspace{-0.4cm}
\caption{\label{f:otherConfigs} (a) Second- and (b) third-preferred configurations for DV near the cores of perfect screw dislocations in 3C-SiC. Carbon and silicon atoms are shown in yellow and blue, respectively. Vacancies are shown as gray spheres. Figures were obtained using OVITO~\cite{0965-0393-18-1-015012}.}
\end{figure*}

In Figs.~\ref{f:otherConfigs}(a) and ~\ref{f:otherConfigs}(b) we show examples of the second- and third-preferred configurations for DV near the cores of perfect screw dislocations in 3C-SiC. The second-preferred configuration, Fig.~\ref{f:otherConfigs}(a), has the silicon vacancy directly at the core and the carbon vacancy siting outside the core. The third-preferred configuration, Fig.~\ref{f:otherConfigs}(b), has both vacancies directly outside the core and in the
second ring from the center.

\subsection{Coordination analysis}

\begin{figure*}[h]
%\vspace{-0.3cm}
%\hspace{2cm}
\includegraphics[scale=1]{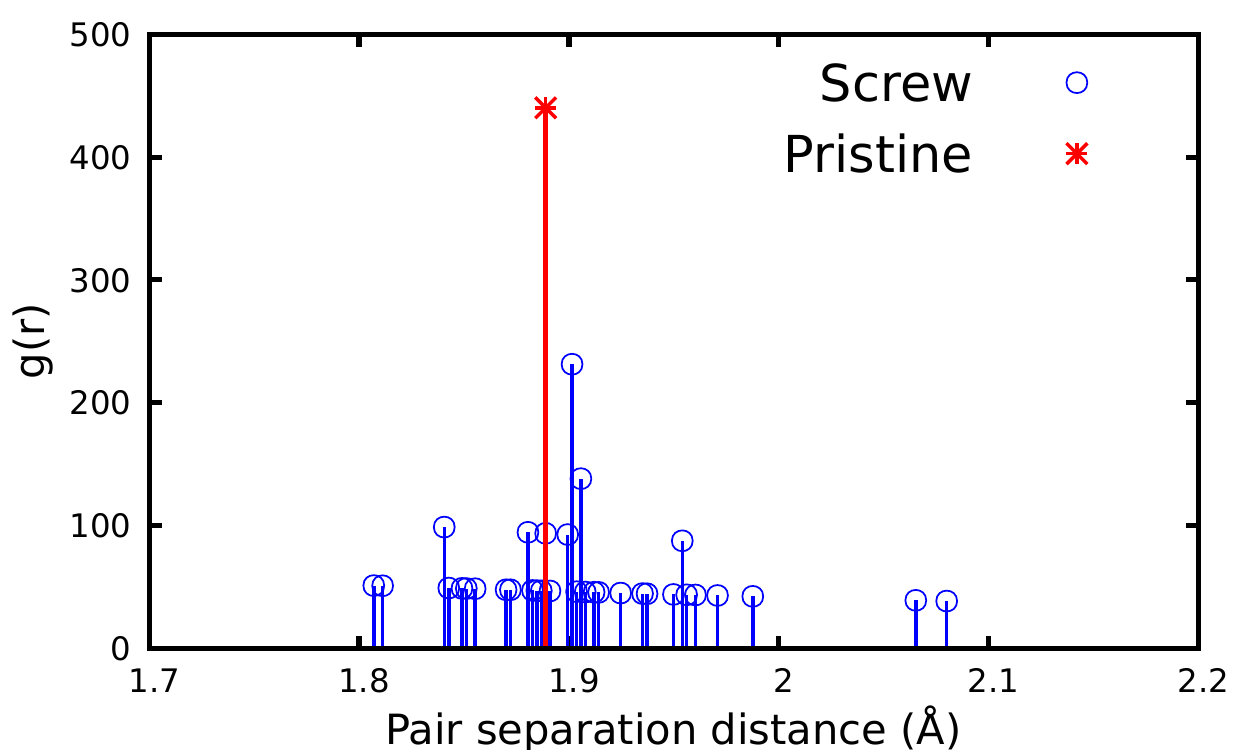} 
%\vspace{-0.4cm}
\caption{\label{f:Coord} coordination analysis of both the pristine 3C-SiC supercell and that of its dislocated counterpart. The analysis was carried out using OVITO~\cite{0965-0393-18-1-015012}.}
\end{figure*}

In Fig.~\ref{f:Coord} we show the radial pair distribution function, $g(r)$, of both the pristine 3C-SiC supercell and that of its dislocated counterpart. As can be seen, the bulk case is characterized by a single nearest neighbour distance. However, once the dislocation dipole is created, the bonding distribution of the supercell exhibits a disorder that explain both the band tail states and the symmetry breaking that lifts the degeneracy of the $e$ states associated to the neutral DV.

\bibliography{Dislos_for_QC}{}

\end{document}